\documentclass[11pt]{elsarticle}
\biboptions{square,comma,sort&compress}
\usepackage[left=1in,right=1in,bottom=1in,top=1in]{geometry}

\makeatletter
\def\ps@pprintTitle{%
	\let\@oddhead\@empty
	\let\@evenhead\@empty
	\def\@oddfoot{}%
	\let\@evenfoot\@oddfoot}
\makeatother

\usepackage[utf8]{inputenc}
\usepackage{fontenc}[t1]

\usepackage{xcolor}

\usepackage{geometry}   % See geometry.pdf to learn the layout options. There are lots.
\usepackage{amssymb}
\usepackage{amsmath}
\usepackage{amsthm}
\usepackage{placeins} %\FloatBarrier
\usepackage{comment}

\usepackage{graphicx}
\usepackage{epstopdf}
\newcommand{\chfnct}{\mathbf{1}}  % Define indicator function
\newcommand{\field}[1]{\mathbb{#1}}

\newcommand {\Q}        {\field{Q} }

\newcommand {\eps}      {\varepsilon}
\theoremstyle{plain}

\newtheorem{prop}{Proposition}[section]

\newtheorem{theorem}[prop]{Theorem}

\newtheorem{cor}[prop]{Corollary}

\newtheorem{conject}[prop]{Conjecture}

\newcommand{\komment}[1]{}

\usepackage[numbers,sort&compress]{natbib}
\begin{document}
\begin{frontmatter} %%NEEDED
	\title{Stochastic Time to Extinction of an SIQS Epidemic Model with Quiescence}
	\author[lab1,lab4]{Usman Sanusi}
    \author[lab1,lab2]{Sona John}
    \author[lab2,lab3]{Johannes Mueller}
    \author[lab1]{Aur\'elien Tellier}
	\address[lab1]{Section of Population Genetics, Department of Life Science Systems, School of Life Sciences, Technical University of Munich, 85354 Freising, Germany} 
	\address[lab2]{Department of Mathematics, School for Computation, Information and Technology, Technical University of Munich, 85748 Garching, Germany} 
	\address[lab3]{Institute for Computational Biology, Helmholtz Center Munich, 85764 Neuherberg, Germany} 
   \address[lab4]{Department of Mathematics and Statistics, Umaru Musa Yar'adua University, Dutsin-Ma Road, P.M.B 2218 Katsina, Katsina State, Nigeria} 
\end{frontmatter} %%NEEDED

\textbf{Abstract}\\
Parasite quiescence is the ability for the pathogen to be inactive, with respect to metabolism and infectiousness, for some amount of time and then become active (infectious) again. The population is thus composed of an inactive proportion, and an active part in which evolution and reproduction takes place. In this paper, we investigate the effect of parasite quiescence on the time to extinction of infectious disease epidemics. We build a Susceptible-Infected-Quiescent-Susceptible (SIQS) epidemiological model. Hereby, host individuals infected by a quiescent parasite strain cannot recover, but are not infectious. We particularly focus on stochastic effects. We show that the quiescent state does not affect the reproduction number, but for a wide range of parameters the model behaves as an SIS model at a slower time scale, given by the fraction of time infected individuals are within the I state (and not in the Q state). This finding, proven using a time scale argument and singular perturbation theory for Markov processes, is illustrated and validated by numerical experiments based on the quasi-steady state distribution. We find here that the result even holds without a distinct time scale separation. Our results highlight the influence of quiescence as a bet-hedging strategy against disease stochastic extinction, and are relevant for predicting infectious disease dynamics in small populations.

%Furthermore, we discover that quiescence affects the quasi-stationary distribution by generating a normal distribution in contrast to the classically expected decreasing distribution observed without quiescence when the basic reproduction number is less than one. 

\section{Introduction}
The general aim of mathematical research on infectious diseases is to contribute towards the eradication and control of epidemics. However, global eradication is hard to achieve. For example, while the eradication of polio had been almost achieved (or thought so), the virus came back into the population because of a decrease of vaccination coverage due to vaccine refusal (and various societal issues) \cite{lahariya2007global}. Furthermore, while global eradication is, indeed, hard to achieve, local extinction of the disease can be observed for some years before the disease re-appears later on in that population. It is also noted that small populations of pathogens are always at higher risk of becoming extinct than larger ones. A typical example of such dynamics is found for the dengue fever incidence in Thailand \cite{cummings2009impact}, where intermittent outbreaks (spikes of incidence) of varying amplitudes are observed over the years. In between outbreaks, some years appear to show almost zero incidence (local possible disease extinction), and the disease re-appears later into the population. Similar observations were made for measles \cite{public2014notifiable, hussain2018anti,grenfell2001travelling, ferrari2008dynamics}. Interestingly, while much epidemiological predictions are based on deterministic mathematical models such as the classic Susceptible-Infected (SI), the stochastic dynamics of local disease extinctions are obviously not well captured by the deterministic approaches. To understand and predict eradication of the disease, it is therefore necessary to model the process by using stochastic processes, namely the birth and death process (of hosts and parasites) under a continuous time and discrete state-space. 
\newline

Parasite quiescence is the ability for the pathogen to be inactive, with respect to metabolism and infectiousness, for some amount of time and then become active (infectious) again \cite{lennon2011microbial,blath2021branching,hurdle2011targeting,blath2020seed, blath2024impact,boots2003population}. This bet-hedging strategy is commonly found in many species of plants, invertebrates and bacteria but also for viral, fungal, and bacterial parasites of animals (including humans) and plants \cite{seger1987bet,sorrell2009evolution}. The parasite population is thus composed of an inactive proportion of individuals, and an active proportion in which evolution and reproduction takes place \cite{balaban2004bacterial,kloehn2021identification, lewis2010persister}. The effect of parasite quiescence on disease epidemiology and epidemics as well as the consequences of quiescence for disease management are yet to be fully understood. The consequences of quiescence can be manifold, especially in models with stochasticity. For example we have shown previously \cite{sanusi2022quiescence} that when two identical parasite strains are present and the transmission rate is high, quiescence periods increase the variability in the number of infected individuals. Conversely, when the transmission rate is low, quiescence reduces this variability. However, when there is competition between parasite strains that have different rates of quiescence, the presence of quiescence leads to a smoothing effect on the fluctuations in the number of infected hosts. This smoothing effect, akin to a moving average, reduces the impact of random variation and ultimately decreases the variability in the number of infected hosts \cite{sanusi2022quiescence}. The efficiency of treatment against infection is also affected by quiescence as bacteria in a quiescent state can be for example resistant to antibiotics \cite{balaban2004bacterial, hurdle2011targeting, lewis2010persister}. Dormancy or quiescence have important consequences for species evolution \cite{tellier2019persistent,lennon2021principles}, by affecting neutral stochastic processes \cite{kaj2001coalescent,blath2021branching,blath2016new} as well as slowing down the process of adaptation \cite{hairston1988rate, koopmann2017, heinrich2018, korfmann2023}. Dormant or quiescent stages constitute a reservoir (storage) of diversity and can re-introduce advantageous genetic variants over time \cite{tellier2019persistent,blath2021branching, shoemaker2018evolution}. Dormancy can therefore evolve as a host or parasite bet-hedging strategy \cite{sorrell2009evolution}, especially as a result of coevolution between hosts and parasites \cite{verin2018host}. The population dynamics of bacteria–virus systems is also affected when the microbial hosts has the ability to transition into a dormant state upon contact with virions, thus evading infection \cite{blath2021virus, blath2023microbial}. Conversely and interestingly, the consequence of the storage effect of quiescent parasites has been less investigated \cite{sorrell2009evolution}. Following our previous results \cite{sanusi2022quiescence}, we hypothesize that parasite quiescence may prevent such stochastic extinction and generate revival of epidemics in a population, by constituting a dormant reservoir. \newline

The main goal of this paper is thus to test this hypothesis. As a reference, we briefly recap the known results for an SIS epidemic model, with the reproduction number, the quasi-steady state distribution (also called Yaglom limit), and the time to extinction as introduced by N{\aa}sell~\cite{naasell2011extinction}. We then incorporate a quiescence phase to the SIS model, so that the extended model is called SIQS, and first show that the reproduction number does not change. Note that a parasite strain is defined as quiescent within an infected host individual when it is inactive (no host mortality as a result of this quiescent infection) and non-infectious (no disease transmission by the quiescent strain). A host individual infected by a quiescent parasite cannot be infected by another parasite (as our model considers only single strain infection epidemiology). By abuse of language, a host individual infected by a quiescent strain can also be referred to as being quiescent, with respect to the epidemiology dynamics and disease transmission. We then derive the equations for the Yaglom limit, which allows us to derive the time to extinction (according to N{\aa}sell's definition) also in this case. As the central finding, we compare the timing of the SIQS model with that of an SIS model, under the assumption that infected individuals rapidly jump between the I and the Q states. Time scale analysis for Markov processes indicate that the SIS and SIQS models behave alike if the time in the SIS model runs slower by a factor given by the proportion of time an infected individual spends in the I state. As the case of individuals rapidly oscillating between I and Q is biologically only of minor relevance, we numerically investigate how far the finding can be stretched, that is, if the finding is still approximately true if the parameters become small. We indeed find that the rapid transitions, though necessary for the mathematical argument chosen, is not necessary for the system. In many cases, the rapid oscillation can be replaced by a large population size: It is central that always a fixed fraction of infected individuals are in the I state. This situation can be reached by a rapid oscillation of single individuals. However, it is also possible that each individual only slowly jumps between the two states, but a large population size ensures a similar effect as rapid oscillations: If we have a large number of infecteds, also then (by the law of large numbers) a fixed fraction of them will actually be infectious. As long as the time scale of the epidemics is slower than the time scale of the infected period, either rapid oscillations or a large population size ensures that the SIQS and the SIS model behave similarly on the respective time scale. In that, the time to extinction is reduced, and in that the Q state indeed prevents the infection to go rapidly extinct.

\section{Model and model analysis}

We start off with an SIS model, which we then extend to an SIQS model; the SIS model will firstly be a reference model 
to put the results of the SIQS model into context. Secondly, we briefly review the ideas about the time to extinction 
developed by N{\aa}sell and studied by others~\cite{naasell2011extinction, allen2010introduction, muller2015methods, wormser2008modeling} for the convenience of the reader in case of this simple model. 

\subsection{SIS model}

We consider a model with fixed size $N$, $S$ ($I$) size of susceptible (infected) class, standard incidence with contact rate $\beta$, and recovery rate $\nu$. 
The deterministic version of the model thus reads
\begin{eqnarray*}
S' &=& -\beta \frac{SI} N + \nu I\\
I' &=& \beta \frac{SI} N - \nu I,
\end{eqnarray*}
with reproduction number $R_0=\beta/\nu$, stationary states $(S,I)=(N,0)$ and $(S,I)=N(1/R_0, 1-1/R_0)$, 
and the usual threshold theorem (for $R_0<1$ the uninfected equilibrium is locally asymptotically -- in the given case 
even globally -- stable).\\
The individual based stochastic version of the model ($I_t\in\{0,..,N\}$ and $N$ are integers, $S_t=N-I_t$), 
$$ I_t\rightarrow I_t+1\quad\mbox{ at rate } \beta (N-I_t)I_t/N.$$
The main challenge with this model is the fact that $I_t=0$ is an absorbing state. The general theory of Markov processes 
indicates that the process almost surely ends up eventually in the uninfected equilibrium, irrespective of the reproduction 
number's value. There are basically two well known ways to deal with this fact: Either the process is coupled to a linear
birth-death process as long as $I_t\ll N$~\cite{ball1995strong}, with the result that major outbreak only is possible with 
probability $1-1/R_0$, that is, with positive probability only if $R_0>1$. The second idea is to acknowledge that the process $I_t$ 
will die out fast if $R_0<1$, but for $R_0>1$ will perform a random walk around the deterministic equilibrium $N(1-1/R_0)$ for a long while before going extinct. The time to extinction is explored. It is this second route we aim to use in the present paper.\par\medskip 

The time to extinction depends in general on the initial state. To circumvent this arbitrariness, Nasell proposes to 
consider the Yaglom limit (the long term behavior of the process conditioned on non-extinction) as the initial state. For $R_0>1$  and $N$ large, the process can be well approximated by an Ornstein-Uhlenbeck process around the endemic equilibrium \cite{andersson2012stochastic,van1995stochastic}, and thus this is a perfectly natural choice.\par\medskip 
In order to obtain the Yaglom limit, we use the master equations: Let $p_i(t)=P(I_t=i)$, then 
\begin{eqnarray}
 \frac d {dt} p_i = -(\beta i(N-i)/N+\nu i)p_i+(\beta i(N-(i-1))/N)\,p_{i-1}+\nu\,(i+1)\,p_{i+1}.
\end{eqnarray}
Next we turn to the process conditioned on non-extinction, $I_t|I_t<0$. If we define $q_i(t)=P(I_t=i|I_t>0)$ for $i=1,..,N$, 
we find
\begin{eqnarray*}
 q_i(t) &=& \frac{P(I_t=i\mbox{ and }I_t>0)}{P(I_t>0)}=\frac{p_i(t)}{1-p_0(t)}\\
 \quad\Rightarrow \quad 
\frac d {dt} q_i &=& [-(\beta i(N-i)/N+\nu i)q_i+(\beta i(N-(i-1))/N)\,q_{i-1}+\nu\,(i+1)\,q_{i+1}]- \nu q_1\, q_i.
\end{eqnarray*}
The quasi-steady state thus satisfies
\begin{eqnarray}\label{eigen}
 [-(\beta i(N-i)/N+\nu i)q_i+(\beta i(N-(i-1))/N)\,q_{i-1}+\nu\,(i+1)\,q_{i+1}] =  \nu q_1\, q_i(t).
\end{eqnarray} 
The time to extinction, according to Nasell, is defined as the arrival time in the state $I_t=0$ if we start with $p_i(0)=q_i$ 
for $i=1,..,N$ and $p_0(0)=0$. Due to eqn.~\eqref{eigen} we have 
$$ p_i(t) = e^{-\nu q_1 t}\, q_i,\qquad p_0(t) = 1-e^{-\nu q_1 t}$$
and thus the time to extinction is exponentially distributed with rate $\mu q_1$.
\begin{theorem}[Nasell] The time to extinction for a population starting in the Yaglom limit 
is exponentially distributed with rate $\nu q_1$, where $q_1$ 
is the probability for the number infecteds is $1$ in the Yaglom limit. 
\end{theorem}
 In~\cite{naasell2011extinction} approximations are discussed which show that this time to extinction strongly increases if $R_0$ crosses the threshold $1$ and the population size is large.

\subsection{SIQS model}
We augment the model by a quiescent state. A person that enter the $I$ state jumps at rate $\rho$ into the state $Q$, and returns
from $Q$ into $I$ at rate $\zeta$. For the deterministic model, we not only have the states $S$ and $I$, but also the population 
density of the quiescent individuals $Q$, 
\begin{eqnarray*}
S' &=& -\beta \frac{SI} N + \nu I\\
I' &=& -\rho I+\zeta Q+\beta \frac{SI} N - \nu I\\
Q' &=& \,\,\,\, \rho I-\zeta Q
\end{eqnarray*}
Also here, the first question is the value of the reproduction number.

\begin{prop}
The reproduction number of the SIQS model is given by $R_0=\beta/\nu$.
\end{prop}
{\bf Proof: } The reproduction number is the spectral radius of the next generation operator; as we have two classes of infected 
individuals ($I$ and $Q$), we need obtain a matrix. We follow~\cite{martcheva2015introduction}, and rewrite the ODE describing the fate of an infected individual as follows
$$ \frac d{dt} \left(\begin{array}{c}I\\Q\end{array}\right)
= \left(\begin{array}{c} \beta \frac{SI} N\\0\end{array}\right)
- \left(\begin{array}{c}\rho I-\zeta Q + \nu I\\-\rho I+\zeta Q\end{array}\right)
= {\mathcal F}-\mathcal{V}$$
and introduce the matrices $F$ (the Jacobian w.r.t. $I,Q$ of ${\mathcal F}$ at $S=N$), and $V$ (the Jacobian w.r.t. $I,Q$ of ${\mathcal V}$), 
$$ F = \left(\begin{array}{cc}
\beta & 0 \\0&0
\end{array}\right),\qquad 
V = \left(\begin{array}{cc}
\rho+\nu & -\zeta  \\
-\rho  & \zeta 
\end{array}\right).
$$
The next generation generator is then given by 
$$ A=FV^{-1} = \left(\begin{array}{cc}
\beta/\nu & \beta/\nu   \\
0  & 0
\end{array}\right)
$$
such that the reproduction number reads $R_0=\rho(A)=\beta/\nu$.\par\qed\par\medskip

\begin{figure}
\centerline{\includegraphics[width=10cm]{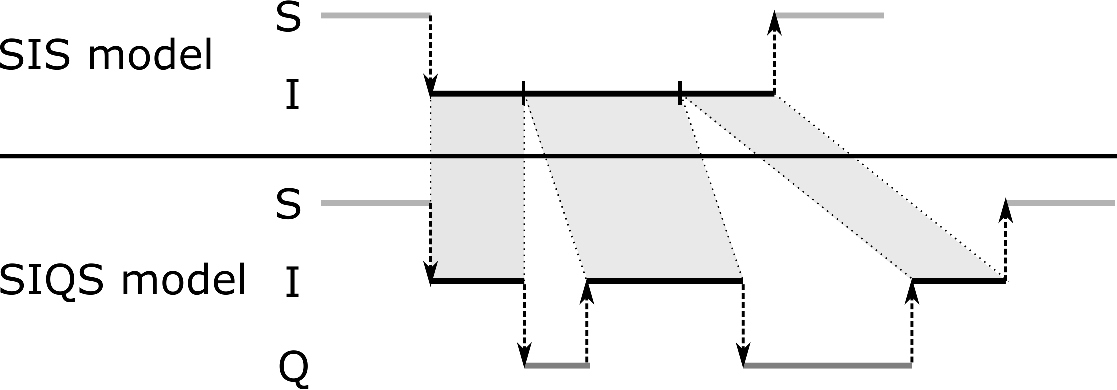}}
\caption{The infectious period of an individual in an SIS model (upper part) and the SIQS model (lower part), where the infected period is divided into several infectious periods, 
interrupted by excursions into the quiescent state. In that, the length of the infected period is extended, 
while the total length of the infectious period does not change.}
\label{sisInfect}
\end{figure}

The corresponding threshold theorem (the uninfected state $(S,I,Q)=(N,0,0)$ is locally asymptotically stable if $R_0<1$, 
and unstable for $R_0>1$) is a consequence of the general theory which can be found in~\cite{martcheva2015introduction}, or can be readily justified
by the usual linearization at the uninfected equilibrium together with the Hartman-Grobman theorem.
\begin{cor} The uninfected equilibrium $(S,I,Q)=(N,0,0)$ is locally asymptotically stable if $R_0<1$, 
and unstable for $R_0>1$.
\end{cor}

It is particularly interesting that the SIS and the SIQS model have an identical reproduction number. At first glance, it can be interpreted as the quiescent state having no influence on the epidemic process. Note that the reproduction number incorporates, however, no information about time -- and, as we will see, it is exactly the timing which is affected by quiescence. If a host individual becomes infected and jumps into the I state, it will go back and forth between states I and Q. While nothing happens in state Q, in state I the individual contributes to infections, and is able to recover. The time the individual spends in I (remove all time intervals of the individual in state Q) is identical to the SIS model (recovery rate is not affected for I-individuals) and it is as infectious as in the SIS model (infection rate is not affected), also see Fig.~\ref{sisInfect}. That is, the excursion into Q delays the effect of an I individual (time points at which this I individual infect others) but does not affect the number of the secondary cases. We will later use exactly this observation to connect the time scale of the SIQS and the SIS model. To better understand the time scales, we now move towards the time to extinction of a stochastic SIQS process.

\par\medskip

For the stochastic version of the model, let $(I_t,Q_t)$ indicates the 
number of active infectious and quiescent individuals, the number of susceptibles is given by $S_t=N-I_t-Q_t$, and 
\begin{eqnarray*}
(I_t,Q_t) &\rightarrow& (I_t+1,Q_t)\quad\quad \,\,\,\, \mbox{ at rate } \beta I_t (N-I_1-Q_t)/N\\
(I_t,Q_t) &\rightarrow& (I_t-1,Q_t)\quad\quad \,\,\,\, \mbox{ at rate } \nu I_t\\
(I_t,Q_t) &\rightarrow& (I_t-1,Q_t+1)\quad \mbox{ at rate } \zeta I_t\\
(I_t,Q_t) &\rightarrow& (I_t+1,Q_t-1)\quad \mbox{ at rate } \rho Q_t. 
\end{eqnarray*}
We again aim at the time to extinction. The way proposed by Nasell carries over: The master equations for $p_{i,j}=P((I_t,Q_t)=(i,j))$
read 
\begin{eqnarray}
\frac d {dt} p_{i,j} &=& -(\beta i(N-i-j)/N+\nu i+\zeta i+\rho j)p_{i,j} 
+ \beta (i-1)(N-(i-1)-j)/N\,p_{i-1,j}\\
&&+ \nu(i+1)\,p_{i+1,j}+\zeta (i+1) p_{i+1,j-1} + \rho (j+1) p_{i-1,j+1}.\nonumber 
\end{eqnarray}
We then introduce the process conditioned on non-extinction, $(I_t,Q_t)|I_t+Q_t>0$. If $q_{i,j} = P((I_t,Q_t)|I_t+Q_t>0)$ for $i=0,..,N$, $j=0,..,N-i$, $(i,j)\not = (0,0)$, we 
obtain in exactly the same way as above the ODE's for $q_{i,j}$, 
\begin{eqnarray}
\frac d {dt} q_{i,j} &=& -(\beta i(N-i-j)/N+\nu i+\zeta i+\rho j)q_{i,j} 
+ \beta (i-1)(N-(i-1)-j)/N\,q_{i-1,j}\\
&&+ \nu(i+1)\,q_{i+1,j}+\zeta (i+1) q_{i+1,j-1} + \rho (j+1) q_{i-1,j+1}- \nu q_{1,0}\, q_{i,j}.\nonumber 
\end{eqnarray}
Therewith we find the Yaglom limit, which then satisfies 
\begin{eqnarray}
\nu q_{1,0}\, q_{i,j} &=& -(\beta i(N-i-j)/N+\nu i+\zeta i+\rho j)q_{i,j} 
+ \beta (i-1)(N-(i-1)-j)/N\,q_{i-1,j}\\
&&+ \nu(i+1)\,q_{i+1,j}+\zeta (i+1) q_{i+1,j-1} + \rho (j+1) q_{i-1,j+1}.\nonumber 
\end{eqnarray}
If we start in the Yaglom limit ($p_{i,j}(0)=q_{i,j}$ for $(i,j)\not = (0,0)$, and $p_{0,0}=0$), we again obtain that 
$$ p_{0,0} = 1-e^{-\nu q_{1,0}}$$
such that the time to extinction is exponentially distributed. 
\begin{cor}
The time to extinction $T$ for the SIQS model, given that we start in the Yaglom limit, is 
exponentially distributed, 
$$ T\sim \mbox{Exp}(\nu q_{1,0}).$$
\end{cor}

A comparison of the timing would be possible if $q_1$ for the Yaglom limit of the SIS model, and $q_{1,0}$ for that of the SIQS model 
would be available. While for the SIS model good approximations are known~\cite{naasell2011extinction}, good estimates are not available for the SIQS model. 
The intuition of course indicates that the probability measure in the SIQS model is distributed to many more states, and hence the 
probability per state is smaller. That is, we expect $q_{1,0}\ll q_1$, and the time scale of the SIQS model to be slower. \par\medskip 

To better understand the difference in the time scale, we introduce a time scale into the SIQS process itself: We assume that the rates 
$\zeta$ and $\rho$ are large, such that we have a fast process (jumping between I and Q) and a slow process (the change in the total number of infected persons). 
We then can use ideas from singular perturbation theory, as discussed in the case of Markov processes by Yin and Zhang~\cite{yin2005discrete}, in order to simplify 
the SIQS process, and in that, to replace the SIQS model by an approximately equivalent SIS model. In this way we are able to compare the time scale of an SIS and an SIQS model. 

\begin{theorem}\label{timescale}
Assume that $\zeta=\hat\zeta/\eps$ and $\rho=\hat\rho/\eps$. For $\eps$ small, the SIQS model $(I_t,Q_t)$ with rate constants $\beta, \nu,\zeta,\rho$ 
behaves as an SIS model $\tilde I_t$ with rate constants $\tilde \beta$, $\tilde \nu$, where
\begin{eqnarray}
\tilde I_t = I_t+Q_t,\qquad \tilde \beta = \frac{\rho}{\rho+\zeta}\,\beta,\qquad  \tilde \nu = \frac{\rho}{\rho+\zeta}\,\nu
\end{eqnarray}
in the sense $I_t+Q_t$ converges in distribution to $\tilde I_t$ for $\eps\rightarrow 0$, after an initial time layer of size ${\cal O}(\eps)$. 
\end{theorem}
\begin{figure}
\centerline{\includegraphics[width=7cm]{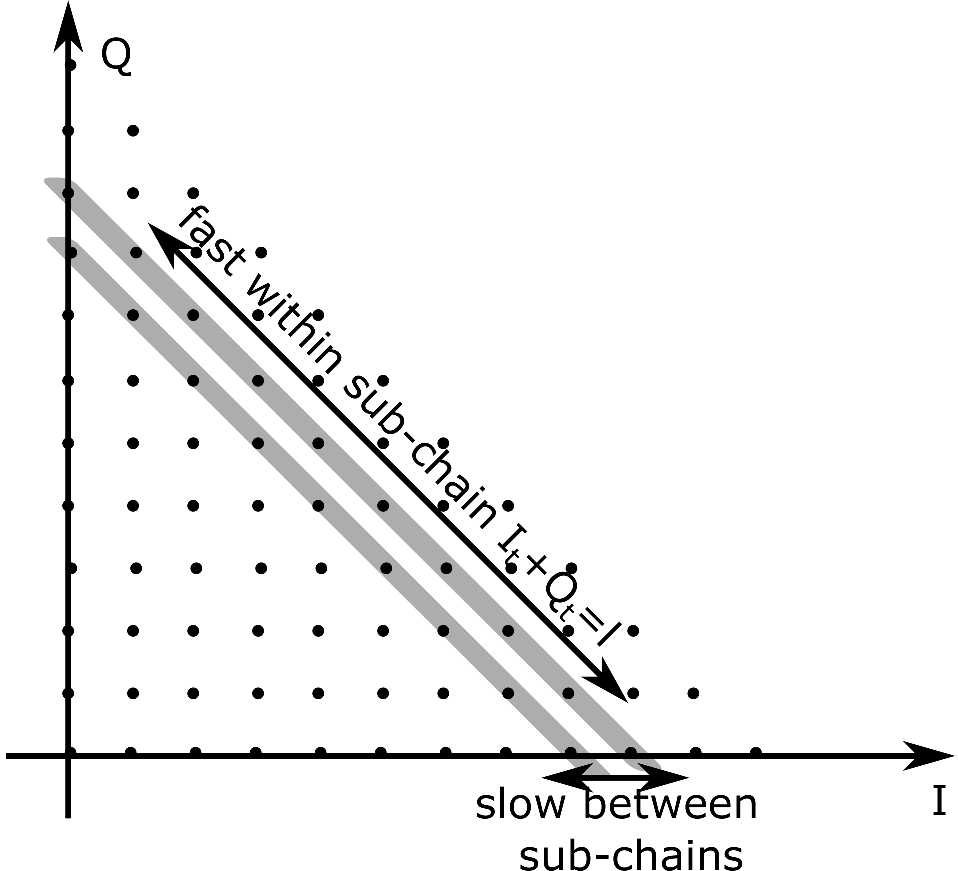}}
\caption{Idea of the proof. Sub-chains with $I_t+Q_t=\ell$ are considered (indicated by gray regions): On the one hand their total probability mass is $r_\ell=P(S_t+Q_t= \ell)$, 
on the other hand the probability of the states within such a chain is $s^{(\ell)}_i=p_{i, \ell-i}$. The dynamics within a sub-chain is fast, such that the sub-chain is in its quasi-equilibrium, 
while the dynamics between the sub-chains is slow, such that the overall dynamics can be reduced to $r_\ell$ only.}\label{ideaProof}
\end{figure}
{\bf Proof: } The idea of the proof is based on the monograph by Yin and Zhang~\cite{yin2005discrete}, who discuss the effects of time scale separation within a Markov process. We expect that 
$I_t+Q_t$ will change only slowly, while infected individuals will rapidly oscillate between I and Q until they recover. This intuition suggests to 
rewrite the master equations for the SIQS model (see Fig.~\ref{ideaProof}), and to introduce $r_\ell(t)=P(S_t+Q_t=\ell)= \sum_{i=0}^\ell p_{i,\ell-i}$. Moreover, 
we change the notation, and move from $p_{i,j}$ to the sub-chain given by $\{(I_t,Q_t)\,:\,I_i+Q_t=\ell\}$ 
with  probabilities $s^{(\ell)}_i(t) = P((I_t,Q_t)=(i,\ell-i))=p_{i,\ell-i}$. Therewith, we write  
(with the understanding that $p_{i,j}=0$ if $(i,j)$ is outside the feasible index combinations $\{(i,j)\,:\, 0\leq i,j, i+j\leq N\}$ 
resp.\ $s^{(\ell)}_i=0$ for $i<0$ or $i>\ell$), 
\begin{eqnarray*}
\frac d {dt} r_\ell &=& 
 -\sum_{i=0}^\ell(\beta i(N-\ell)/N+\nu i+\hat \zeta i/\eps+\hat \rho(\ell-j)/\eps)p_{i,\ell-i} \\
&&+ \sum_{i=0}^\ell \beta (i-1)(N-(i-1)-\ell-i)/N\,p_{i-1,\ell-i}\\
&&+ \sum_{i=0}^\ell \nu(i+1)\,p_{i+1, \ell-j}+\sum_{i=0}^\ell\zeta (i+1) p_{i+1,\ell-i-1}/\eps + \sum_{i=0}^\ell\hat \rho (\ell-i+1) p_{i-1,\ell-i+1}/\eps\\
&=& 
 -\sum_{i=0}^\ell(\beta i(N-\ell)/N+\nu i)s^{(\ell)}_i 
+ \sum_{i=0}^\ell \frac \beta N (i-1)(N-(\ell-1))\,s^{(\ell-1)}_{i-1} + \sum_{i=0}^\ell \nu(i+1)\,s^{(\ell+1)}_{i+1}\\
&=& 
 - (\beta (N-\ell)/N+\nu )\sum_{i=0}^\ell i s^{(\ell)}_i 
+ \frac \beta N (N-(\ell-1))\sum_{i=0}^{\ell-1} i\,s^{(\ell-1)}_{i} + \nu\,\sum_{i=0}^{\ell+1} i\,s^{(\ell+1)}_{i}.
\end{eqnarray*}
We find indeed that the fast time scales do not explicitly appear in the equation for $r_\ell$. If, however, we focus on the probability-flux of the sub-chain $s^{(\ell)}_i$, we find 
\begin{eqnarray*}
\eps \frac d {dt} s^{(\ell)}_i 
&=&
 -(\eps \beta i(N-\ell)/N+\eps \nu i+\hat \zeta i+\hat \rho (\ell-i))s^{(\ell)}_i
+ \eps \beta (i-1)(N-(i-1)-j)\,s^{(\ell-1)}_{i-1}/N\\
&&+ \eps \nu(i+1)\,s^{(\ell+1}_{i+1}+\hat \zeta (i+1) s^{(\ell)}_{i+1} + \hat \rho (\ell-i+1) s^{(\ell)}_{i-1}.
\end{eqnarray*} 
Thus the Fenichel theory~\cite{o1991singular} indicates that, for $\eps$ small, after an ${\cal O}(\eps)$ initial time layer, 
the solution for $s^{(\ell)}_i$ settles down on the slow manifold, given by (take $\eps\rightarrow 0$)
$$  
0
=
 -(\hat \zeta i+\hat \rho (\ell-i))s^{(\ell)}_i+ \hat \zeta (i+1) s^{(\ell)}_{i+1} + \hat \rho (\ell-i+1) s^{(\ell)}_{i-1}.
$$
We aim to reduce the master equations to this slow manifold. Thereto, we determine $r_\ell\sum_{i=0}^\ell i s^{(\ell)}_i/r_\ell$, where 
$\sum_{i=0}^\ell i s^{(\ell)}_i/r_\ell=E(I|I+Q=\ell)$. Heuristically, it is clear that $E(I|I+Q=\ell)=\ell\rho/(\rho+\zeta)$, as each infected 
individual rapidly and independently oscillates between I and Q where the probability for the I-state is $\rho/(\rho+\zeta)$. More formally, 
by means of the  detailed balance equation, it is $\hat \zeta i s^{(\ell)}_i = \hat\rho (\ell-i+1) s^{(\ell)}_{i-1}$, and 
$$ s^{(\ell)}_i = \frac{\hat\rho (\ell-i+1)}{\hat \zeta i}s^{(\ell)}_{i-1} = \prod_{k=1}^i\frac{\hat \rho (\ell-k+1)}{\hat \zeta k}s^{(\ell)}_{0}.$$
Herein $s^{(\ell)}_0(t)$ is determined by $\sum_{i=0}^\ell s^{(\ell)}_i(t)=r_\ell(t)$, 
$$ s^{(\ell)}_0 = \frac{r_\ell}{\sum_{i=0}^\ell \prod_{k=1}^i\frac{\hat \rho (\ell-k+1)}{\hat \zeta k}}.$$
With $\prod_{k=1}^i\frac{ (\ell-k+1)}{ k} = \frac{\ell!}{(\ell-i)!i!}={\ell\choose i}$ we obtain
\begin{eqnarray*} 
 \sum_{i=0}^\ell i\, s^{(\ell)}_i 
&=&
r_\ell\,\, \frac
{ \sum_{i=0}^\ell i\prod_{k=1}^i\frac{\hat \rho (\ell-k+1)}{\hat \zeta k}} 
{\sum_{i=0}^\ell \prod_{k=1}^i\frac{\hat \rho (\ell-k+1)}{\hat \zeta k}}
= 
r_\ell\,\, \frac
{ \sum_{i=0}^\ell i\ x^i {\ell\choose i}} 
{\sum_{i=0}^\ell x^i {\ell\choose i}}\bigg|_{x=\hat\rho/\hat\zeta}\\
&=& r_\ell\,\,\frac {x \frac d{dx} (1+x)^\ell}{(1+x)^\ell}\bigg|_{x=\hat\rho/\hat\zeta}
= r_\ell\,\,\frac{\ell\,\hat\rho/\hat\zeta}{1+\hat\rho/\hat\zeta}.
\end{eqnarray*} 
Therewith, the ODE for $r_\ell(t)$ becomes 
\begin{eqnarray*}
\frac d {dt} r_\ell 
&=& 
 - (\tilde \beta \ell (N-\ell)/N+\tilde \nu\ell )r_\ell 
+ \tilde\beta (\ell-1)\,(N-(\ell-1)) r_{\ell-1}/N + (\ell+1)\tilde \nu\,\,r_{\ell+1}.
\end{eqnarray*}
That is, $r_\ell$ satisfies approximately ($\eps$ small) the master equations for the process $\tilde I_t$, which proves the theorem.\par\qed\par\medskip

In the reference SIS model, $\beta$ and $\nu$ are scaled by $\rho/(\rho+\zeta)$, which can be thought of as time simply running slower. We have the following corollary.

\begin{figure}
\begin{center}
(a)\rotatebox{270}{\includegraphics[width=6cm]{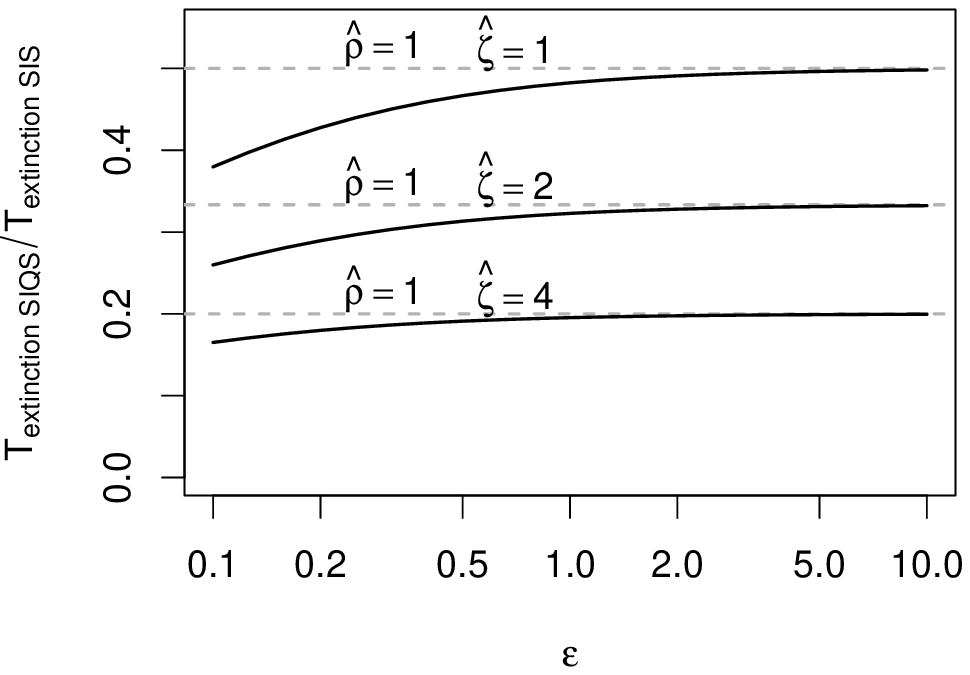}}(b)\rotatebox{270}{\includegraphics[width=6cm]{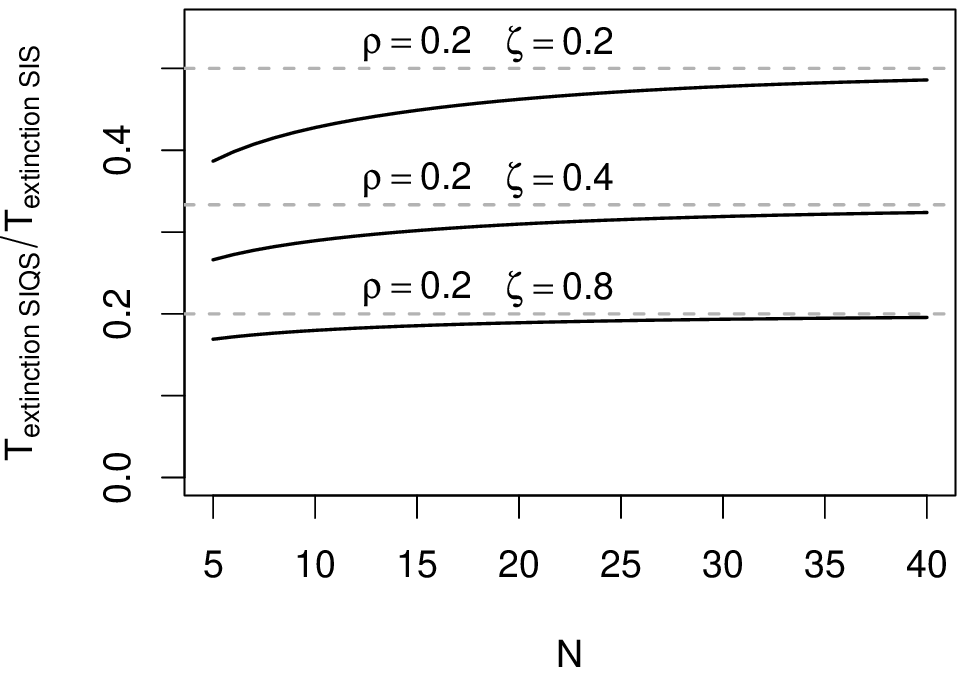}}
\end{center}
\caption{Quotient of the time to extinction of the SIQS model over that of the SIS model, 
numerically determined based on the quasi-steady state distributions (solid lines) together with the expected ratio $\frac{\rho}{\rho+\zeta}$ in case of large rates (dashed gray line). We choose $\beta=1$, $\nu=0.5$. (a) Effect of the time scale: $\zeta=\eps\hat\zeta$, $\rho=\eps\hat\rho$, $\eps$ as indicated on the $x$-axis, $\hat\zeta$ and $\hat \rho$ as indicated in the plot, $N=10$. 
(b) Effect of the population size $N$: $\rho$ and $\zeta$ as indicated in the plot, $N=5,..,40$, as indicated on the $x$-axis.}\label{sim1}
\end{figure}

\begin{cor}\label{timeExtinct}
Let $T_{sis}$ denote the time distribution to extinction for an SIS model with rate constants $\beta$, $\nu$, and 
$T_{siqs}$ the time to extinction for an SIQS model with rate constants $\beta,\nu,\rho,\zeta$. If $\rho,\zeta\gg 1$ while $\beta,\nu={\cal O}(1)$, then approximately 
$$ T_{siqs} = \frac{\rho}{\rho+\zeta}\, T_{sis}.$$
\end{cor}
As expected, in contrast to the reproduction number, the timing of the SIQS model -- here the time to extinction -- is affected by quiescence. Numerical analysis of the time to extinction (Fig.~\ref{sim1}(a)) confirms that for $\zeta$,$\nu$ large, the relation 
$ T_{siqs} = \frac{\rho}{\rho+\zeta}\, T_{sis}$ is an appropriate approximation, while it becomes invalid if the rates are small.\par\medskip 

However, we note that only the ratio of $\rho/\zeta$ enters in the result. This is because the invariant measure (which again is time-invariant) in the oscillation between the I and the Q state is of importance for the dynamics, while the absolute time span an individual spends in Q is not (that) central. The actual values of $\rho$ and $\zeta$ 
influence the switching time scale of a single individual, but do not centrally affect the time course of the epidemic at the population level.\par\medskip 

This insight, interesting in itself, also might imply that the result carries over to situations where the time scale separation of theorem~\ref{timescale} and corollary~\ref{timeExtinct} 
is not given. For example, for a large population a certain fraction of $\{I,\,Q\}$-individuals are in the I state. In effect, it is not each single infected individual, but the total amount of infected individuals that guarantees that the fraction $\frac{\rho}{\rho+\zeta}$ of infected individuals are infectious. However, as every newly
infected person starts and ends its infectious period in the I-state, we suggest that the I-state might be underestimated by the theory introduced above, especially in the case of small rates (only few oscillations). 
We re-discuss the fraction of time spend in I by an infected host individual, particularly for the case that 
the rates $\nu$, $\rho$ and $\zeta$ are of the same order. The infectious period of an individual follows the equation (recall the definition of matrix $V$ above) 
$$ \frac d {d\tau} \left(\begin{array}{c} p_I(\tau)\\p_Q(\tau)\end{array}\right) 
=  \left(\begin{array}{cc} -(\zeta+\nu) & \rho\\ \zeta & -\rho \end{array}\right) 
\left(\begin{array}{c} p_I(\tau)\\p_Q(\tau)\end{array}\right) = -V \left(\begin{array}{c} p_I(\tau)\\p_Q(\tau)\end{array}\right), \quad 
\left(\begin{array}{c} p_I(0)\\p_Q(0)\end{array}\right)  = \left(\begin{array}{c} 1\\0\end{array}\right)  = e_1.
$$ 
An newly infected individual starts in the I state. The expected time that an individual spends in the I state before recovery is given 
by $E(\int \chfnct(\mbox{her state at time $t$ is } I)\, dt) = \int p_I(t)\, dt$. 
Similarly, the expected time an infected host individual harbours a quiescent parasite is given by $\int p_Q(t)\, dt$. 
Therewith, 
$$ 
\left(\begin{array}{c} \mbox{time in I}\\  \mbox{time in Q}\end{array}\right) 
=\left(\begin{array}{c} \int_0^\infty p_I(\tau)\,d\tau \\ \int_0^\infty p_Q(\tau)\,d\tau \end{array}\right) 
= \int_0^\infty e^{-V \tau}\, e_1\,d\tau  = V^{-1}e_1 
= \frac 1 {\rho\nu} \left(\begin{array}{c} \rho \\  \zeta \end{array}\right)
. 
$$
The proportion of time spent in $I$ is therefore still $\rho/(\rho+\zeta)$, regardless of the apparent advantage of the I-state (we start and end 
the infectious period in the I-state) and the size of the parameters (no time scale separation required). 
\begin{conject} We expect the result of theorem~\ref{timescale} and corollary~\ref{timeExtinct} to be stable if the population size $N$ is large, even if the time scale separation is not given, 
as long as the prevalence of infection does not change much on the time scale given by 
an infectious excursion of an individual. 
\end{conject} 
This conjecture is supported by the numerical analysis in Fig.~\ref{sim1}(b).

\section{Conclusion}
We develop here a two-dimensional structure called the SIQS model (Susceptible-Infectious-Quiescent-Susceptible) by expanding the analysis of a stochastic SIS (Susceptible-Infectious-Susceptible) model to add a quiescence phase. The SIS model is frequently used to explain the transmission of infectious diseases in a community where people might contract the disease, recover, and then become once more vulnerable. In situations when specific microparasites (bacteria, viruses or fungi) can enter a dormant or quiescent state, rendering them less sensitive to therapy and immune responses, the addition of the quiescence phase to build SIQS model enables a more thorough understanding of the epidemiological dynamics. Some parasite strains can undergo quiescence, a state of inactivity or dormancy, under specific circumstances, such as when they are exposed to drugs, unfavourable surroundings or viruses  \cite{lewis2000programmed, sat2001programmed, blath2021virus, blath2023microbial}. We examine how quiescence affects the dynamics of the infectious disease using both analytical and computational methods. \\
We show in the present study, that quiescence mainly affects the time scale of an epidemic, if individuals only infect and recover in the I but not in the Q state. Particularly, the reproduction number (which does not include information about the time scale) is not affected. We are able to prove this result in case that infected individuals rapidly oscillate between the I and the Q state, but show -- by means of numerical simulations -- that a rapid oscillation is not necessary for the result to hold; specifically if the population size (and thus also the number of infected hosts) is sufficiently high. Please note that the relative prevalence does not need to be large, as our finding seemingly only depends on the absolute number of infected host individuals. The central mechanism relies on a fixed proportion of those infected to actually be infectious. It does not seem to matter if this fixed proportion is achieved by 1) each single individual rapidly jumping between I and Q, or 2) the fluctuations in the fraction of infectious individuals averaging out due to the large number of infected persons. In any case, the influence of the Q state reduces the infection and recovery rates in the same way. \\
We note that the assumption of a homogeneously mixing population is, generally, problematic for a large population size. Basically, any infection runs along some contact graph. We could disregard the contact graph using time scale arguments. Heuristically, if the infection chains connect any individual with any other individual within a time span at which the prevalence hardly changes, the contact graph does not play a central role. For large populations and sparse graphs (e.g., when the population is organized in patches of individuals), this assumption might become violated/problematic. Further realistic effects may be important and affect the reliability of SIS types of models, such as a seasonal change of parameters (school holidays, rainy/dry seasons and alike) or coevolution between hosts and parasites. In these cases, quiescent parasites may harbour advantage and have more chances to persist \cite{sorrell2009evolution, verin2018host} and the SIQS model be more realistic. Our findings, indeed, suggest that quiescence serves as a bet-hedging strategy against potential extinction, stabilising the parasite population over time and lowering the risk that the disease would go extinct. \\

Our study sheds light on the intricate relationship between disease dynamics and parasite quiescence. Quiescence can have a considerable impact on disease control and management strategy by altering the stability, quasi-stationary distribution, and time to extinction of the disease. The decision-makers can more accurately forecast and prepare for the duration and effects of epidemics by taking the quiescence period into account in infectious disease models. This knowledge can inform the development of targeted interventions to control the spread of infectious diseases and ultimately improve public health outcomes.

\newpage
\bibliographystyle{plain}
\bibliography{source.bib}

\end{document}